\documentclass[12pt]{article}
\newlength{\vshift}
\usepackage{graphicx}
\newlength{\hshift}

\usepackage{amssymb}
\usepackage{amsmath}
\usepackage{amsthm}
\usepackage{amsopn}
\def\beq{\begin{equation}}
\def\eeq{\end{equation}}
\def\bea{\begin{eqnarray}}
\def\eea{\end{eqnarray}}

\title{Regularizing future cosmological singularities with varying speed of light}
\author{F. Shojai$^{1,2}$, A. Shojai$^{1,2}$, M. Sanati$^1$\\ $^1$Department of Physics, University of Tehran,\\$^2$Foundations of Physics Group, School of Physics,\\ Institute for Research in Fundamental Sciences (IPM),\\Tehran, Iran.}
\date{}
\begin{document}
\maketitle
\begin{abstract}
Cosmological models may result in future singularities. We show that, in the framework of dynamical varying speed of light theories, it is possible to regularize those singularities.  
\end{abstract}
\section{Introduction}
The current observational data, interpreted  in the framework of FLRW cosmology, indicates that the dominant content of the universe is in the form of dark energy whose equation of state parameter, $w$ is close to $-1$. It is quite possible that dark energy scenarios, drive the universe towards some type of singularity. Fore example, in a simple case with a time independent phantom equation of state parameter ($w<-1$), the universe evolves towards a big-rip singularity \cite{caldwell}. This kind of singularity is an example of other new exotic singularities that may happen in the future of the universe. 
These singularities usually violate some of the energy conditions. They can be classified according to their properties. This classification of the future cosmological singularities is summarized in the next section.

An important question is that is it possible to find a mechanism to change the nature of these singularities or remove them? This question was recently investigated in the framework of varying physical constants theories \cite{dab, dabmar}. Some examples are varying gravitational constant \cite{brans}, varying electron charge \cite{bek} and  varying speed of light (VSL) \cite{mof,alb,bar}. The latter plays an important role in the special and general relativity. A naive introduction of variable speed of light breaks Lorentz invariance. There would be a preferred frame, in which Einstein’s equations are valid in which the time dependent speed of light inserted as input. But it can be shown \cite{mag} that by introducing a new time-like coordinate,  one can retrieve local Lorentz invariance and general covariance. VSL theories can be motivated by solving the problems of standard cosmological model. It is shown that a larger speed of light in the early universe can solve the horizon problem \cite{mof, alb, bar}.

In this paper we investigate how a VSL theory can regularize some of the future finite-time  cosmological singularities \cite{dab}. We use a general scalar tensor action in which the dynamical speed of light is coupled to gravity non-minimally. Then we shall find the necessary coupling and the potential function to regularize some of the future singularities. 

\section{Future Singularities}
As stated before, one of the features of a dark energy dominated universe is the possibility of appearance of future exotic singularities, leading to the violation of energy conditions. For our later convenience we review the classification of such singularities \cite{cop, nojiri, ski}:

\begin{itemize}

\item Type I (Big Rip Singularity): For $t \rightarrow{t_ s}$ , $a \rightarrow {\infty}$, $\rho \rightarrow {\infty}$ and $|p| \rightarrow {\infty}$.\\
This singularity happens in a phantom dark energy model with an equation of state parameter $w < -1$ \cite{cald} and all the energy conditions are violated.

\item Type II (Sudden Future Singularity): For $t \rightarrow{t_ s}$ , $a \rightarrow {a_s}$, $\rho \rightarrow {\rho_s}$ and $|p| \rightarrow {\infty}$.\\
This singularity was first investigated by Barrow \cite{barr}. He constructed a
class of models which leads to sudden future singularity for which the weak and strong energy conditions hold but the dominant energy condition is violated. This is a pressure singularity that keeps the density, the scale factor and the Hubble parameter finite.

\item Type III (Finite Scale Factor or Big Freeze Singularity): For $t \rightarrow{t_ s}$ , $a \rightarrow {a_s}$, $\rho \rightarrow {\infty}$ and $|p| \rightarrow {\infty}$.\\
This singularity has been discovered in the model of \cite{nojir} and then found in \cite{lopez} for phantom models with generalized Chaplygin gas.

\item Type IV  (Big Brake or Big Separation Singularity): For $t \rightarrow{t_ s}$ , $a \rightarrow {a_s}$, $\rho \rightarrow {0}$, $|p| \rightarrow {0}$ but  derivatives of the Hubble parameter may diverge.\\
Such a singularity can be found in the tachyonic cosmological model \cite{gor}.

\item Generalized Sudden Future Singularities: For $t \rightarrow{t_ s}$ $a \rightarrow {a_s}$, $\rho \rightarrow {\rho_s}$,  $|p| \rightarrow {p_s}$ and the  derivatives of the pressure may diverge \cite{ba}. For this singularity, all the energy conditions are satisfied.

\item $w$-singularity: For $t \rightarrow{t_ s}$ , $a \rightarrow {a_s}$, $\rho \rightarrow {0}$, $|p| \rightarrow {0}$, the equation of state parameter diverges while the derivatives of the Hubble
parameter do not. The energy conditions do not seem to violate at this singularity \cite{dabro}. 

Moreover in \cite{cat} a number of different finite-time singularities are studied using generalized power series expansion of the scale factor. The energy conditions are analyzed in the vicinity of these events. For a $w$ singularity the scale factor admits a Taylor series in which the linear and quadratic terms are absent \cite{jamb}. 

From a different point of view, the authors of \cite{jam} discuss the behaviour of geodesics in the presence of a variety of cosmological singularities. In this way they are able to classify singularities into two groups, weak and strong. In the former case the space-time is geodesically complete and  in the latter is incomplete.
\end{itemize}
\section{Regularization of Future Singularities in Varying $c$ Models}
In order to investigate the possibility of regularization of future singularities within VSL theories, we use the scale factor recently proposed by Dabrowski \cite{dab}. It admits 
a variety of singularities by adjusting the range of some parameters. It is given by:
\begin{equation}
a(t)=a_s\left(\frac{t}{t_s}\right)^m \exp{\left(1-\frac{t}{t_s}\right)}^n
\label{a}
\end{equation}
where $m$ and $n$ are arbitrary constants. $t_s$ and $a_s$ are the finite cosmic time and the scale factor at which a particular singularity may occur depending on the values of  $m$ and $n$. By this form of the scale factor, the universe have started with a Big-Bang singularity at $t=0$ and eventually can reach to another singularity at the future. According to \cite{dab}, the possible future singularities of the above cosmological model can be classified as it is shown in the table (\ref{table1}).
\begin{table}
\begin{center}
\begin{tabular}{c|c|c}
$n$ & $m$ & Singularity \\
\hline
\hline 
$(1,2)$ & arbitrary & Sudden Future (Type II)\\
\hline
arbitrary & $( -\infty,0 )$ & Big Rip (type I) \\
\hline
$(0,1)$ & arbitrary &  Finite Scale Factor (type III) \\
\hline
$n>2$ & $0$ & $w$ \\
\end{tabular}
\end{center}
\caption{Future singularities of the cosmological model given by equation (\ref{a})}
\label{table1}
\end{table}

It is quiet acceptable to believe that time variation of physical constants, especially the speed of light,  may regularize some of the cosmological singularities. There are two ways of making a physical constant time dependent. The naive way is to feed time dependence of any physical constant as an input function in the equations of motion of the theory. Thus the basic postulate of this method is that the equations of motion are valid with variable constants but this can only be true in one frame. This method proposed in \cite{mof, alb, bar},  as a solution to some of the problems of the standard cosmological model. Recently , it is shown that this way of forcing the speed of light or the gravitational constant to vary,  can regularize some of the future singularities \cite{dab,dabmar,dabden}. In fact in these papers, appropriate time dependent constants are suggested such that some of the singularities, types II and III, are regularized.

To have a physically meaningful regularization, it should be covariant. Therefore the regularization should be done in dynamically variable constant theories. Here we are interested to regularize the type  II, III and $w$ singularities by VSL. 

As Ellis \cite{ell} pointed out, to change a constant into a dynamical variable, one needs to start from a new Lagrangian in which the constant is replaced by a  dynamical field with the corresponding dynamical terms.  One of the simplest actions suitable for a dynamical speed of light is that of the scalar--tensor model. In Jordan picture it is \cite{esp}:

\begin{equation}
S={1\over 16\pi G} \int d^4x \sqrt{-g} \Bigl(F(\Psi)R - 2U(\Psi) -
~g^{\mu\nu}
\partial_{\mu}\Psi
\partial_{\nu}\Psi
 \Bigr)
+ S_m[\phi_i,g_{\mu\nu}] \label{1}
\end{equation}
in which we have to assume $F(\Psi)=(c/c_0)^4$ and $ d^4x=dt  d^3x$ to have the correct limit of constant speed of light.  $U(\Psi)$ is the potential energy of the scalar field  $\Psi$, $c_0$ is the constant velocity of light and hereafter we shall put $8\pi G=c_0^4=1$. $S_m$ is the action of the matter fields, $\phi_i$, and doesn't involve the $\Psi$ field, that is to say, we have assumed that the matter is minimally coupled to gravity. 

Variation of
the action (\ref{1}) with respect to the metric and the scalar field gives:
\[
F(\Psi) (R_{\mu\nu}-{1\over2}g_{\mu\nu}R) = T_{\mu\nu}^{matter}
+ \nabla_\mu\Psi\nabla_\nu\Psi - {1\over 2}g_{\mu\nu}
(\nabla_\alpha\Psi)^2
\]
\begin{equation}
+\nabla_\mu\nabla_\nu F(\Psi) - g_{\mu\nu}\nabla_\mu\nabla^\mu
F(\Psi) - g_{\mu\nu} U(\Psi)
\label{2}
\end{equation}
\begin{equation}
\nabla_\alpha\nabla^\alpha \Psi = - \frac{1}{2}{dF\over d\Psi}R
+{dU\over d\Psi}\ \label{si}
\end{equation}
and variation with respect to the matter fields gives the matter equations of motion.

Applying the field equations (\ref{2})-(\ref{si}) to FLRW universe, we get:
\begin{equation}
3F(H^2+\frac{k}{a^2}) = \rho +{1\over 2} \dot{\Psi}^{2} - 3 H
\dot{F} + U \label{H2}
\end{equation}
\begin{equation}
-2F(\dot{H}-\frac{k}{a^2}) = (\rho+p) + \dot{\Psi}^{2} + \ddot{F}
- H\dot{F} \label{dotH}
\end{equation}
\begin{equation}
(\ddot{\Psi}+3H\dot{\Psi}) = 3{dF\over d\Psi}\left(\dot{H} + 2
H^2+\frac{k}{a^2} \right) -{dU\over d\Psi} , \label{ddotPhi}
\end{equation}
The first two equations are VSL-FLRW equations and the other is the equation of motion of 
$\Psi$ field. $H$, $\rho$  and $p$ are the Hubble parameter, energy and pressure densities of the matter field and dot denotes derivative with
respect to the time--like coordinate $t$. 
From (\ref{a}), the Hubble parameter and it's derivative are:
\begin{equation}
H(t)=\frac{m}{t}-\frac{n}{t_s}\left (1-\frac{t}{t_s}\right )^{n-1}
\end{equation}
\begin{equation}
\dot{H}(t)=-\frac{m}{t^2}+\frac{n(n-1)}{t_s^2}\left (1-\frac{t}{t_s}\right )^{n-2}
\end{equation}
In order to regularize the singularities, let's assume that the time dependence of the coupling function $F$ and the field $\Psi$  are given by:
\begin{equation}
F=F_0\left(1-\frac{t}{t_s}\right)^\beta
\label{f}
\end{equation}
\begin{equation}
\Psi=\Psi_s \left(\frac{t}{t_s}\right)^\alpha+\Psi_0\left(1-\frac{t}{t_s}\right)^\gamma
\label{ps}
\end{equation}
where $F_0$, $\Psi_s$, $\Psi_0$, $\beta$, $\alpha$ and $\gamma$ are arbitrary constants. By substituting (\ref{f}) and (\ref{ps}) into (\ref{H2}) and (\ref{dotH}), we find that:
\[
\rho(t)=3F_0  \left [\frac{m^2}{t^2}\left(1-\frac{t}{t_s}\right)^\beta +\frac{n^2}{t_s ^2}\left(1-\frac{t}{t_s}\right)^{\beta+2(n-1)}
-\frac{2mn}{t t_s}\left(1-\frac{t}{t_s}\right)^{\beta+n-1}\right]
\]
\[
-\frac{\alpha^2\Psi_s^2}{2t_s^2}\left(\frac{t}{t_s}\right)^{2(\alpha-1)}-\frac{\Psi_0^2\gamma^2}{2t_s^2}\left(1-\frac{t}{t_s}\right)^{2(\gamma-1)}+
\frac{\Psi_0\Psi_s\alpha\gamma}{t_s^2}\left(\frac{t}{t_s}\right)^{\alpha-1}\left(1-\frac{t}{t_s}\right)^{\gamma-1}
\]
\begin{equation}
-\frac{3F_0\beta}{t_s}\left[\frac{m}{t}\left(1-\frac{t}{t_s}\right)^{\beta-1}-\frac{n}{t_s}\left(1-\frac{t}{t_s}\right)^{\beta+n-2}\right]-U(t)
\label{ro}
\end{equation}

\[
P(t)=F_0\left[\frac{m(2-3m)}{t^2}\left(1-\frac{t}{t_s}\right)^{\beta}-
\frac{2n(n+\beta-1)}{t_s^2}\left(1-\frac{t}{t_s}\right)^{\beta+n-2}
-\frac{3n^2}{t_s^2}\left(1-\frac{t}{t_s}\right)^{\beta+2(n-1)}\right .
\]
\[
\left . +\frac{6mn}{t_s^2}\left(1-\frac{t}{t_s}\right)^{\beta+n-1}
+\frac{2\beta m}{t t_s}\left(1-\frac{t}{t_s}\right)^{\beta-1}
-\frac{\beta(\beta-1)}{t_s^2}\left(1-\frac{t}{t_s}\right)^{\beta-2}\right]
\]
\[
-\frac{1}{2t_s^2}\left[\alpha^2\Psi_s^2\left(\frac{t}{t_s}\right)^{2(\alpha-1)}+\Psi_0^2\gamma^2\left(1-\frac{t}{t_s}\right)^{2(\gamma-1)}-
2\Psi_0\Psi_s\alpha\gamma \left(\frac{t}{t_s}\right)^{\alpha-1}\left(1-\frac{t}{t_s}\right)^{\gamma-1}\right]
\]
\begin{equation}
+U(t)
\label{p}
\end{equation}
Since for a sudden future singularity $1<n<2$, in order to get finite values of the density and pressure it is required that $\beta\geq 2$ and $\gamma\geq 1$ and the potential term contains the positive powers of $(1-\frac{t}{t_s})$. Keeping this in mind, let's to find the appropriate form of potential from (\ref{ddotPhi}) to regularize sudden future singularity. Multiplying (\ref{ddotPhi}) by $\dot{\Psi}$, the time derivative of the potential is easily found as follows:
\[
\dot U(t)=
\frac{3F_0\beta m(1-2m)}{t^2 t_s}\left(1-\frac{t}{t_s}\right)^{\beta-1}+
\frac{\Psi_0\Psi_s\alpha\gamma}{t_s^2}\left[\frac{\alpha-1}{t_s}+\frac{6mt_s}{t^2}\right]\left(\frac{t}{t_s}\right)^{\alpha-2}\left(1-\frac{t}{t_s}\right)^{\gamma-1}
\]
\[
-\frac{3F_0\beta}{t_s}\left[\frac{n(n-1)}{t_s^2}\left(1-\frac{t}{t_s}\right)^{n+\beta-3}
+\frac{2n^2}{t_s^2}\left(1-\frac{t}{t_s}\right)^{2n+\beta-3}
-\frac{4mn}{t}\left(1-\frac{t}{t_s}\right)^{\beta+n-2}\right]
\]
\[
-\frac{1}{t_s^3}\left[\Psi_s^2\alpha^2(\alpha-1)\left(\frac{t}{t_s}\right)^{2\alpha-3}
+\Psi_s\Psi_0\alpha\gamma(\gamma-1)\left(\frac{t}{t_s}\right)^{\alpha-1}\left(1-\frac{t}{t_s}\right)^{\gamma-2}
-\Psi_0^2\gamma^2(\gamma-1)\left(1-\frac{t}{t_s}\right)^{2\gamma-3}\right]
\]
\[
-\frac{3}{t_s^2}\left[\frac{\Psi_s^2\alpha^2 m}{t}\left(\frac{t}{t_s}\right)^{2(\alpha-1)}
+\frac{m\Psi_0^2\gamma^2}{t}\left(1-\frac{t}{t_s}\right)^{2(\gamma-1)}
-\frac{n\Psi_s^2\alpha^2}{t_s}\left(\frac{t}{t_s}\right)^{2(\alpha-1)}\left(1-\frac{t}{t_s}\right)^{n-1}
\right .
\]
\begin{equation}
\left .
-\frac{n\Psi_0^2\gamma^2}{t_s}\left(1-\frac{t}{t_s}\right)^{2\gamma+n-3}
+\frac{2\Psi_s\Psi_0\alpha\gamma n}{t_s}\left(\frac{t}{t_s}\right)^{\alpha-1}\left(1-\frac{t}{t_s}\right)^{n+\gamma-2}\right]
\label{dotu}
\end{equation}
At the vicinity of $t_s$, integrating the dominant terms of the above expression gives:
\begin{equation}
U(t)=U_0+ \left \{ 
\begin{matrix}
\frac{\Psi_s^2\alpha^2(\alpha-1+3m)}{t_s^2} \left(1-\frac{t}{t_s}\right)& \textrm{if } \gamma>2 \\
\frac{\Psi_s\Psi_0\alpha\gamma}{t_s^2} \left(1-\frac{t}{t_s}\right)^{\gamma-1}& \textrm{if } \gamma<2
\end{matrix}
\right .
\label{usfs}
\end{equation}
where $U_0$ is an integration constant. In this limit the energy and pressure densities are:
\begin{equation}
\rho(t)=-\left(U_0+\frac{\Psi_s^2 \alpha^2}{2 t_s^2}\right)+\frac{\Psi^2\alpha^2(1-6m)}{2 t_s^2}\left(1-\frac{t}{t_s}\right)
\label{ro1}
\end{equation}
\begin{equation}
P(t)=\left(U_0-\frac{\Psi_s^2 \alpha^2}{2 t_s^2}\right)+\left \{ 
\begin{matrix}
\frac{-F_0\beta (\beta-1)}{t_s^2}\left(1-\frac{t}{t_s}\right)^{\beta-2}& \textrm{if } \beta<1+\gamma, \beta<3 \\
\frac{2\Psi_0\Psi_s \alpha\gamma}{t_s^2}\left(1-\frac{t}{t_s}\right)^{\gamma-1}& \textrm{if } \beta>1+\gamma, \gamma<2 \\
\frac{\Psi_s^2\alpha^2(\alpha-1+3m)}{t_s^2} \left(1-\frac{t}{t_s}\right)& \textrm{if } \beta>3, \gamma>2 
\end{matrix}
\right.
\label{p1}
\end{equation}
From (\ref{ro1}) and (\ref{p1}) one can find that for $\beta\geq 2$ and $\gamma\geq 1$, the energy density and pressure vanish at $t_s$. Moreover sudden future singularity is regularized and the equation of state parameter of matter is in this limit is:
\begin{equation}
w=-\frac{U_0-\frac{\Psi_s^2 \alpha^2}{2 t_s^2}}{U_0+\frac{\Psi_s^2 \alpha^2}{2 t_s^2}}
\label{w1}
\end{equation}
which is finite. The varying speed of light is $c_0F_0^{1/4}\left(1-\frac{t}{t_s}\right)^{\beta}$ which also approaches to zero in the vicinity of $t_s$. This is in accordance to the result of \cite{dab}. Moreover slowing and stopping of light is also predicted in loop quantum cosmology \cite{cail}. From (\ref{ps}), one can conclude that for $\gamma>1$, at $t\rightarrow t_s$:
\begin{equation}
\alpha\left(1-\frac{t}{t_s}\right)=1-\frac{\psi}{\psi_s}
\end{equation} 
Substituting this in (\ref{usfs}), the $\psi$-dependence of coupling and potential function is resulted.

One can repeat all the above discussion for other singularities. We will not pursue this calculation here and only report the results. For a finite scale factor singularity, the potential, density and pressure functions are finite if $\beta\geq 2$ and $\gamma\geq 1$. But since $0<n<1$,  at the vicinity of $t_s$:
\begin{equation}
U(t)=U_0+ \left \{ 
\begin{matrix}
\frac{3 F_0\beta n(n-1)}{t_s^2 (n+\beta-2} \left(1-\frac{t}{t_s}\right)^{n+\beta-2} & \textrm{if } n+\beta<\gamma+1 \\
\frac{\Psi_s\Psi_0\alpha\gamma}{t_s^2} \left(1-\frac{t}{t_s}\right)^{\gamma-1}& \textrm{if } n+\beta>\gamma+1
\end{matrix}
\right .
\label{ufsf}
\end{equation}
\begin{equation}
\rho(t)=-\left(U_0+\frac{\Psi_s^2 \alpha^2}{2 t_s^2}\right)+\frac{3F_0\beta n(\beta-1)}{t_s^2 (n+\beta-2}\left(1-\frac{t}{t_s}\right)^{n+\beta-2} 
\label{ro2}
\end{equation}
and $P(t)$ is given by (\ref{p1}).

Also for $w$ singularity, $m=0$ and $n>2$. In this case the potential and density functions are given by (\ref{usfs}) and (\ref{ro1}) with $m=0$ respectively. 
\section{Conclusion}
We have considered a general scalar-tensor theory in which the varying speed of light is a function of the scalar field. Applying this theory to the Friedmann cosmology, we have shown that one can regularize some kinds of singularities including sudden future, finite scale factor and $w$-singularities. To do this, we have assumed a simple time dependence for the coupling function and the scalar field. We have found that sudden future, finite scale factor and $w$-singularities can be regularized by a varying $c(t)$ which is zero at the vicinity of these singularities. It is interesting to note that using the obtained forms of density and pressure in the vicinity of regularized singularity, one can simply observe that the energy conditions are broken.

\textbf{Acknowledgement} This work is  supported by a grant from university of Tehran.


\begin{thebibliography}{99}
\bibitem{caldwell}
R. R. Caldwell, M. Kamionkowski, and N. N. Weinberg, \textit{Phys. Rev. Lett.}, \textbf{91}, 071301, 2003.
\bibitem{dab}
M. P. Dabrowski and K. Marosek, \textit{JCAP}, \textbf{02}, 012, 2013. 
\bibitem{dabmar}
M. P. Dabrowski, K. Marosek and A. Balcerzak,	arXiv:1308.5462,
Proceedings of the Sesto Conference on Varying fundamental constants and dynamical dark energy, Sesto, Italy, July 8-13, 2013.
\bibitem{brans}
C. Brans and R. Dicke, \textit{Phys. Rev.}, \textbf{124}, 925, 1961.
\bibitem{bek}
J. D. Bekenstein, \textit{Phys. Rev. D}, \textbf{25}, 1527, 1982.
\bibitem{mof}
J.W. Moffat, \textit{Int. J. Mod. Phys. D}, \textbf{2}, 351, 1993.
\bibitem{alb}
A. Albrecht and J. Magueijo, \textit{Phys. Rev. D}, \textbf{59}, 043516, 1999.
\bibitem{bar}
J.D. Barrow, \textit{Phys. Rev. D}, \textbf{59}, 043515, 1999.
\bibitem{mag}
J. Magueijo, \textit{Phys. Rev. D}, \textbf{62}, 103521, 2000.
\bibitem{cop}
E. J. Copeland, M. Sami, and S. Tsujikawa, \textit{Int. J. Mod. Phys. D}, \textbf{15}, 1753, 2006.
\bibitem{nojiri}
S. Nojiri, S. D. Odintsov, and S. Tsujikawa, \textit{Phys. Rev. D}, \textbf{71}, 063004, 2005.
\bibitem{ski}
M. P. Dabrowski and T. Denkiewicz, \textit{AIP Conf. Proc.}, \textbf{1241}, 561, 2010.
\bibitem{cald}
R. R. Caldwell, \textit{Phys. Lett. B},\textbf{545},23, 2002.
\bibitem{barr}
J. D. Barrow, \textit{Class. Quant. Grav.}, \textbf{21}, L79, 2004.
\bibitem{nojir}
S. Nojiri and S. D. Odintsov, \textit{Phys. Rev. D}, \textbf{70}, 103522, 2004.
\bibitem{lopez}
M. Bouhmadi-Lopez, P. F. Gonzalez-Diaz, and P. Martin-Moruno, \textit{Phys. Lett. B}, \textbf{659}, 1, 2008.
\bibitem{gor}
V. Gorini, A. Yu. Kamenshchik, U. Moschella, and V. Pasquier,
 \textit{Phys. Rev., D}, \textbf{69}, 123512, 2004.
\bibitem{ba}
J. D. Barrow and C. G. Tsagas, \textit{Class. Quant. Grav.}, \textbf{22}, 1563, 2005.
\bibitem{dabro}
M. P. Dabrowski and T. Denkieiwcz, \textit{Phys. Rev.D}, \textbf{79}, 063521, 2009.
\bibitem{cat}
C. Cattoen and M. Visser, \textit{Class. Quant. Grav.}, \textbf{22}, 4913, 2005.
\bibitem{jamb}
L. Fernandez-Jambrina, \textit{Phys. Rev. D}, \textbf{82}, 124004, 2010.
\bibitem{jam}
L. Fernandez-Jambrina and R. Lazkoz, \textit{Phys. Rev. D}, \textbf{74}, 064030, 2006.
\bibitem{dabden}
M. P. Dabrowski, T. Denkiewicz, C. J. A. P. Martins and P.E. Vielzeuf, arXiv:1406.1007.
\bibitem{ell}
G.F.R. Ellis and J.P. Uzan, \textit{Am. J. Phys.}, \textbf{73}, 3, 240, 2005.
\bibitem{esp}
G. Esposito-Farese and D. polarski, \textit{Phys. Rev. D},
\textbf{63}, 063504, 2001.\\
A. Riazuelo and J. P. Uzan, \textit{Phys. Rev. D}, \textbf{66}, 023525, 2002.
\bibitem{cail}
T. Cailleteau, J. Mielczarek, A. Barrau and J. Grain, \textit{Class. Quantum. Grav.} \textbf{29}, 095010, 2012.

\end{thebibliography}
\end{document}